# A User-Based Authentication and DoS Mitigation Scheme for Wearable Wireless Body Sensor Networks

Nombulelo Zulu
*Department of Information Technology*
*Tshwane University of Technology*
Pretoria, South Africa
nombulelozuluh@gmail.com

Deon P. du Plessis
*Department of Information Technology*
*Tshwane University of Technology*
Pretoria, South Africa
DuPlessisDP@tut.ac.za

Topside E. Mathonsi
*Department of Information Technology*
*Tshwane University of Technology*
Pretoria, South Africa
MathonsiTE@tut.ac.za

Tshimangazdo M. Tshilongamulenzhe
*Department of Information Technology*
*Tshwane University of Technology*
Pretoria, South Africa
TshilongamulenzheTM@tut.ac.za

*Abstract*—Wireless Body Sensor Networks (WBSNs) is one of the greatest growing technology for sensing and performing various tasks. The information transmitted in the WBSNs is vulnerable to cyber-attacks, therefore security is very important. Denial of Service (DoS) attacks are considered one of the major threats against WBSNs security. In DoS attacks, an adversary targets to degrade and shut down the efficient use of the network and disrupt the services in the network causing them inaccessible to its intended users. If sensitive information of patients in WBSNs, such as the medical history is accessed by unauthorized users, the patient may suffer much more than the disease itself, it may result in loss of life. This paper proposes a User-Based authentication scheme to mitigate DoS attacks in WBSNs. A five-phase User-Based authentication DoS mitigation scheme for WBSNs is designed by integrating Elliptic Curve Cryptography (ECC) with Rivest Cipher 4 (RC4) to ensure a strong authentication process that will only allow authorized users to access nodes on WBSNs.

**Keywords**—*User-Based authentication, Denial of Service (DoS) attacks, wireless body sensor networks (WBSNs), Rivest Cipher 4 (RC4), Elliptic Curve Cryptography (ECC)*

## 1. INTRODUCTION

Authentication is a crucial security requirement to avoid attacks against secure communication and to mitigate Denial of Service (DoS) attacks exploiting the limited resources of sensor nodes. According to [11] with the help of Wireless Body Sensor Networks (WBSNs), extensive, and uninterrupted health monitoring becomes easy and possible, and the user can enjoy the medical services provided by the application without limitations. *Alsubaie, Al-Akhras & Alzahrami* [9] DoS attacks are one of the most dangerous forms of attacks that can threaten WBSNs security. In DoS attacks, an adversary gets connected with the WBSNs applications. Once the adversary gets a frequency slot in WBSNs, it starts sending a huge amount of traffic to the sensor node, which makes it impossible for the sensor node to process the traffic coming from legitimate nodes.

Most wireless network users require consistent and secure systems, however, this is not always the case with WBSNs applications, as the communication channels are visible to many attackers due to their unattended deployment nature. Since the information transmitted via the sensor node is very sensitive and critical, user authentication of such information is of prime importance. If sensitive information, such as the medical history is accessed by unauthorized adversaries, the patient may suffer much more than the disease itself, it may result in a wrong prescription and put the life of a patient at risk which can lead to death. Malicious users can access WBSNs' private network and launch DoS attacks, which can breach the availability of medical devices

This paper proposed a User-Based authentication scheme to protect WBSNs by ensuring that data is transferred over the network is kept safe and accessed by only authorized personnel only. The scheme is designed in a way that the user will initiate the process by attempting to login into the network. The first phase in the authentication process is to initialize the server to generate the master key that will be used in the encryption/decryption phase, and the registration of the user will take place to ensure that the user trying to access the network is from the trusted list. The second phase is login which will grant the user access if known/exist else access will be denied, following will be the DoS mitigation step that will ensure that the network has enough power to transport data obtained via sensor nodes. The encryption/decryption process takes place lastly to ensure that the user can access data and view using the master key generated from the initializing step, else deny access if the key used does not match.

The contributions of this paper are as follows: 1) Initializing the server and registration of sensor nodes during data transmission can be used to ensure that data received is from a known and trusted source to minimize the amount of traffic coming from legitimate nodes and mitigate DoS attacks. 2) The use of integrated Elliptic curve cryptography (ECC) with the Rivest Cipher 4 (RC4) method is mainly because the proposed algorithm should be efficient enough and fast when authenticating WBSN users. To make sure only authorized users can have access to the sensitive patient's data. 3) To use User-Based authentication DoS mitigation scheme that will prevent malicious users from connecting to the network and prevent attackers from launching a DOS attack in WBSN applications.

This paper is structured as follows: In section 1 the paper introduced the background of wireless body sensor network applications by outlining the aim, the issue the study is attempting to solve, and the contributions of the paper. In section 2, the study highlighted the work done by the other researchers in WBSN applications. Section 3 presented the workflow of the proposed User-Based aauthentications scheme. In Section 4, this paper discussed the implementation of the proposed scheme. In Section 5, the paper discussed the chosen simulation tool namely MATLAB. In addition, the section presented the simulation results of the compared algorithms namely, User-Based authentication scheme, biometric authentication algorithm, and cryptographic techniques scheme. Lastly, Section 6 concludes the paper and presents future works.

## 2. RELATED WORKS

Over the past years, several studies have been conducted on authentication schemes to protect the information of WBSNs users during the data transmission process.

*Zhang, Wang, Vasilakos & Fang* [21] proposed the Received Signal Strength Indication (RSSI)-based key agreement scheme. Their study showed that the key agreement process is treated as a simplified error-correcting encoding and decoding with only check symbols being passed from the initiating side to the responding side. Their study achieved both security and efficiency. Their study highlighted one big issue being that RSSI-based key generation and agreement have limited data density. The User-Based authentication and DoS mitigation scheme proposed in this paper can achieve both efficiency and security by integrating Elliptic Curve Cryptography (ECC) with Rivest Cipher 4 (RC4). This was achieved by generating the public key from the private key while protecting the private key from being known when disclosing the public key during the user-authentication process which will mitigate DoS from having slots in the network.

The study conducted by [21] presented an overview of attacks, principles, and solutions on the anonymity of two-factor authentication schemes. To improve the current schemes from being stuck with the security-usability tension. They proposed integrating "honeywords", traditionally the purview of system security, with a "fuzzy-verifier". Their scheme solves two issues, it does not only eliminate the security-usability conflict that is considered intractable in the literature, but also achieves security guarantees beyond the conventional optimal security bound. The scheme proposed by [21] resolved the various issues arising from user corruption and server compromise. Although their proposed scheme is computationally efficient, secure, and lightweight, it does not consider the anonymity of identities and unlikability service. During the authentication process, the proposed scheme in this paper can ensure that both anonymity identity and unlikability are achieved.

*Das, Chatterjee & Sing* [16] proposed a biometric-based authentication scheme for hierarchical WBANs, where biometric verification with password verification was used. The efficient cryptographic hash function and symmetric key encryption and decryption algorithm was only required in their scheme and the scheme was lightweight. Later, the study conducted by [22] proved that the scheme could not provide real user anonymity and easy to trace. Their study proposed and used a new anonymous authentication scheme with bilinear pairing for WBANs which proved that the study by [12] was easy to trace. The scheme proposed in this paper provides password verification during the login phase. Encryption/Decryption algorithm is used to ensure that sensitive data is protected during the user authentication process.

The study conducted by [19] proposed a collection of optimisation techniques to reduce the energy cost of watchdog utilisation when maintaining the security of the network at an appropriate level. It includes the theoretical analyses along with the practical algorithms which can schedule the several tasks of the watchdog based on the position of the node and the trustworthiness of the destination nodes. The study conducted by [6], stated and proved that the scheme proposed in the study conducted by [19] does not provide mutual authentication and prone to jamming attacks. The proposed algorithm in this paper, provides mutual authentication by comparing the user logging details with the existing information stored in the system. The system administration will ensure that the user is from the trusted list.

*Chiou, Ying & Liu* [13] proposed a scheme that guarantees anonymity, unlikability, and message authentication for uses. Their scheme has four phases: (1) Healthcare center uploading phase (HUP), (2) Patient uploading phase (PUP), (3) Treatment phase (TP), and (4) Checking phase (CP). The hash function they used was Secure Hashing Algorithm (SHA)-256, the symmetric encryption algorithm was Advanced Encryption Standard (AES), and the signature algorithm is ECC. The scheme allows patients to consult with doctors directly and remotely in a safe way. However, the scheme still fails to provide patient anonymity and message authentication and it only stores patient medical data, without allowing patients to directly access medical advice. The User-Based authentication and DoS mitigation scheme proposed in this paper can provide message authentication.

*Li, Lee, & Weng* [14] presented participant authentication in mobile emergency medical care systems for patient supervision. They propose a secure cloud-assisted architecture for accessing and monitoring health in WBANs. Chaotic maps-based authentication and key agreement mechanisms are utilized to provide data security and mutual authentication. Based on the proposed scheme by [14] designed another dynamic identity and chaotic maps-based authentication scheme and a secure data protection approach to prevent illegal intrusions for medical systems. They also proposed an improved secure authentication and data encryption scheme for smart devices in medical systems. Their scheme exposes the patient and the doctor to the flaw of private key reveal the problem and is failing to provide real-time monitoring service and non-repudiation evidence in doctor diagnosis. The proposed algorithm in this paper can provide a key management mechanism during the encryption/decryption process in order to ensure that the text is encrypted or decrypted using the master key that is only known to the sender.

*Wu, Xu, Kumari & Li* [8] presented a secure three-factor user authentication scheme for Wireless Sensor Network (WSN). However, in 2019, *Mo & Chen* [3] demonstrated that if the user inputs an incorrect password at the login phase in [8] scheme the smartcard does not check whether the password is verified, and the protocol will proceed until Gateway Node (GWN) finds that the login request of the user was invalid, so the GWN performs unnecessary computational resources. In the algorithm proposed in this paper, the authentication process ensures that when the user login details does not exist or unknown and if the credentials provided are incorrect, the user will not be granted access to the system without doing unnecessary computational. Thus, the proposed algorithm produced better network throughput.

*Koya & Deepthi* [7] proposed an anonymous hybrid mutual authentication and key agreement scheme based on Electrocardiography (ECG). However, their scheme is vulnerable to sensor node capture attacks and lacks forward secrecy. Besides, physiological signal-based authentication schemes typically require sensors to monitor unique physiological signals, such as ECG, which makes such schemes lack universality. The integration of ECC and RC4 in this paper ensures that the secrecy in WBSNs during data transmission is maintained.

*Wang & Park* [20] used different mechanisms and consider sensor node types to avoid traffic analysis attacks in WBSNs applications. Set of the delays in different periods to avoid the traffic was added, by using the scheme the optimal solution was obtained to protect the network with no overheads. Their study also highlighted the drawback of this method is that the delay is twice the actual traffic signal length. The proposed scheme in this paper seeks to mitigate DoS attacks in WBSNs that cause traffic in a network. The proposed User-Based authentication scheme in this paper reduced the delay in transmitting data caused by the traffic signal length.

According to [18] ECC based authentication and key agreement protocol for Telecare Medicine Information Systems (TMIS) was proposed and declared that it could provide better security and performance. The same protocol was demonstrated in the study conducted by [17] and identified that the protocol is vulnerable to a replay attack and has an inefficient authentication phase. This paper integrates ECC and RC4 in order to ensure that the user authentication phase is efficient.

*Gope & Hwang* [15] proposed an efficient mutual authentication and key agreement scheme for global mobility networks. However, the study conducted by [8] pointed out that [15], scheme lacks session key update and wrong password detection mechanisms, lacks forward secrecy, and is vulnerable to a DoS attack. This paper proposed a User-Based authentication scheme that mitigates DoS attacks in WBSNs.

The review of literature has indicated that previous studies did not propose any scheme that can mitigate DoS attacks efficiently in the WBSNs. To overcome the identified limitation, this paper proposed an efficient User-Based authentication and DoS mitigation scheme. The proposed scheme was designed by integrating ECC and RC4 in order to obtain a strong encryption and decryption process during the data transmission process in WBSNs.

## 3. USER-BASED AUTHENTICATION SCHEME DESIGN

The User-Based authentication and DoS mitigation scheme is designed by integrating ECC and RC4 in order to ensure that sensitive data is kept safe within the WBSNs. The proposed scheme has four (4) phases namely initialization and registration, login, DoS mitigation, and encryption/decryption. Based on the review of the literature, the attacker may try to change the information by positioning itself next to the victim. However, this cannot be the case in the proposed scheme since the DoS mitigation step will ensure that the patient's data is protected by preventing such attacks.

### 3.1 User-Based Authentication scheme flowchart

The overview of the entire system design is demonstrated as shown in Fig 1. The flowchart illustrates the logic behind carrying out all four phases in the User-Based authentication and DoS mitigation scheme.

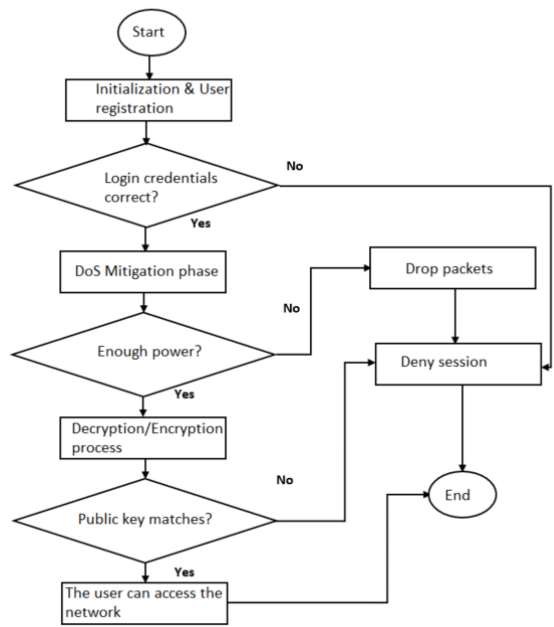

Fig 1: User-Based authentication scheme flowchart

ECC and RC4 were integrated in order to create faster, smaller, and more efficient cryptographic keys. According to *Roy & Khatwani* [10] ECC is a widely-used technique in multi-factor authentication. ECC generates keys through the properties of the elliptic curve equation instead of the traditional method of generation as the product of very large prime numbers. ECC was used in this paper, because it helps to establish equivalent security with lower computing power and battery resource usage [23]. RC4 was used in this paper for the encryption/decryption phase in order to protect patient's data [10]. RC4 was chosen due to its low computational requirements, speed, and simplicity when implemented.

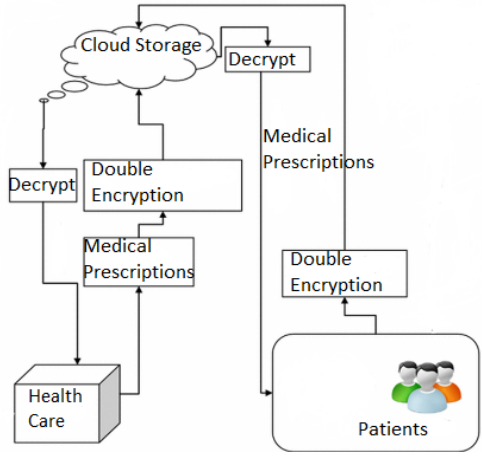

Fig 2: ECC algorithm in WBSNs

The flow of information from patients (sensor node) to a healthcare server and vice versa is shown in Fig 2.

Step 1 – Patients: Information is gathered by inserting a sensor node in a patient, that will read the information of a certain disease a patient is treated for. The initializing phase takes place in this step where the system administrator generates a master key that will be used by the healthcare server to decrypt the message.

Step 2 – Double encryption: this is where the patient's information gets encrypted by using the encryption algorithm to ensure that only authorised users can view the information being sent.

Step 3 – Cloud Storage: Information gathered from the patient via sensor nodes is saved in the cloud storage while waiting to be accessed by the healthcare institution, in this step the DoS mitigation algorithm is executed to make sure that data is protected against attacks while sitting in the cloud storage.

Step 4 – Decrypt: Data/information is decrypted by the registered user using the shared master key only known to the receiver and the healthcare institution can be to access and read the patient's record. Registration and Logic phases are executed in this step to validate that the user logging in is registered.

Step 5 – Healthcare: The assigned doctor reads the patient's data and provides the necessary prescription which is transfer using the same process. Data gets encrypted and the information is saved in the cloud storage while waiting to be transferred to the sensor node by the system administrator. Information is accessible using the master key that is only known to the receiver.

*3.2 The Developed User-Based Authentication scheme*

The four phases are carried out, where all of them are executed one after another in order to ensure that patient's sensitive data is protected, and their privacy is maintained. If the response received from the server was negative the session was denied, and the user was requested to retry to enter the correct login details or register first in the system first if the user does not exist. Below is the proposed User-Based authentication scheme designed by integrating ECC and RC4 in order to minimize the security issues in WBSNs.

### User-Based Authentication Algorithm

Initialization phase:
1. Choose $r$ and Generate $k_{Ser}$

Registration phase:
2. Choose $ID_{Sn}$, $r = P_{ks}$.
3. Compute $r = A_{sn} \| B_{Sn} \| B_{Sn} \| ID_{Sn}$
4. $AP's \rightarrow ID_{AP}$ through secure channel

Authentication phase:
Step 1
1. $A$ sends $B$ the $P_{ks}$
2. $B$ sends $A$ public key $P_{ks}$
3. For each authentication do
4. Generate $ID_{Sn}$; $S1$; $S2$; $t1$

Step 2:
Forward $(A_{sn}; S1; S2; t1; ID_{AP})$.to the server
5. $m$ Identity $= ID_{AP}$

Step 3:
6. Generate $n1$ and $ti$
7. Compute $S1 = B_{Sn} + n1$
8. $A_{sn}|\# = A_{sn} \leftrightarrow B_{Sn} +$
9. $A_{sn}| \#(t_{new} - t1)$

Step 4:
10. Check if $A_{sn}, S1, ID_{AP}$ are present in the database
11. Validate $t1$
12. Compute $N2^*$
13. Sensor validation $= A_{SN}^+$; $B_{Sn}^+$; $P_{ks}^+$

DoS mitigation phase:
14. Read all $m$
15. Matches the $ID_{Gw}$
16. $ID_u = Enc(ID_u + ID_{Gw})$ compute
17. If
18. Power is more/equals to SN power
19. Send packets
20. Else
21. Drop packets

Encryption phase:
22. Verify if $m$ is received
23. For $i \leftarrow 0$ to $255$
24. $m = a * b$
25. Data is encrypted resulting in $C$
26. $C = m(mod\ n)$
27. Computed $(S[i], S[j]) = C(XOK\ S[k]$

Decryption phase
28. Extract $k$ from value $J$
29. $sk = j \otimes J.i$
30. Extract $k$ from value $S$
31. $S[k] = H(sk \| k)$
32. Decrypt $m = J\ (mod\ i)$
33. Checks $sk* = s$
34. END

### 4. PROPOSED SCHEME IMPLEMENTATION

The implementation of a User-Based authentication scheme is presented in the section. The scheme is implemented using IEEE 802.11s model developed using MATLAB version 2020a on the computing platform, a machine running Windows 10 with 8GB memory was used. 100 sensor nodes were randomly arranged with a radius of 20, and the nodes distribution is randomly distributed and clustered. The connection radius is 1 unit, and the minimum distance between nodes is set to 0.6 cm because setting mesh routers closer to each other is not common in the real-world. MATLAB was chosen in this paer because of its high-performance language for technical computing [2]. MATLAB integrates computation, visualization, and programming in an easy-to-use environment where real-life problems and solutions are expressed.

MATLAB is written in C language and it supports the cross-platform operating system [12]. In addition, MATLAB provides various command-line functions as well as tools for measuring, analysis, and visualization of data like digital filtering, signal processing, and modulation of signalling. MATLAB and Simulink are integrated together, which allows the researchers to amend the functions in the toolbox, and simulator [1]. The simulator is also useful when generating waveforms and building to test systems. MATLAB provides several applications to scientists, engineers, researchers, and educators for technical computing.

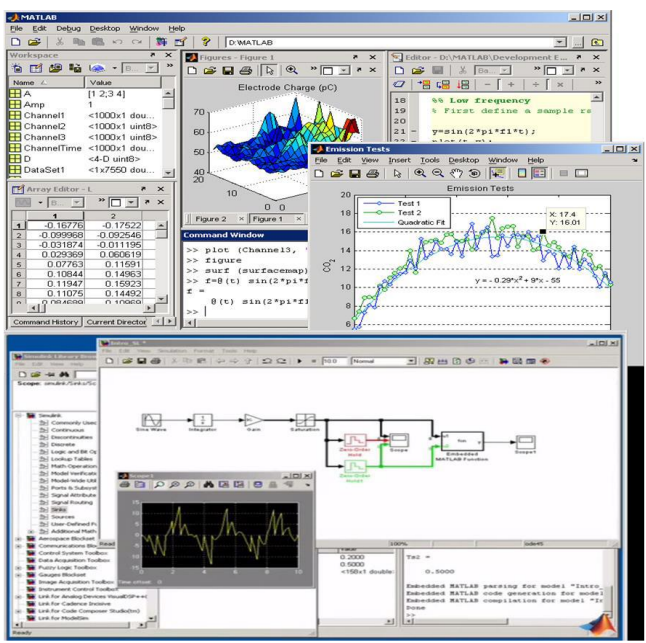

Fig 3: MATLAB simulator interface

### 4.1 Simulation Set-up

The main purpose of simulating the User-Based authentication algorithm was to evaluate if the proposed algorithm brings the improvement in the WBSNs aplications. MATLAB is a very high-performance simulation tool, the latest version currently available for download is MATLAB 2021a. The configured simulation topology in MATLAB is shown in Fig. 4.

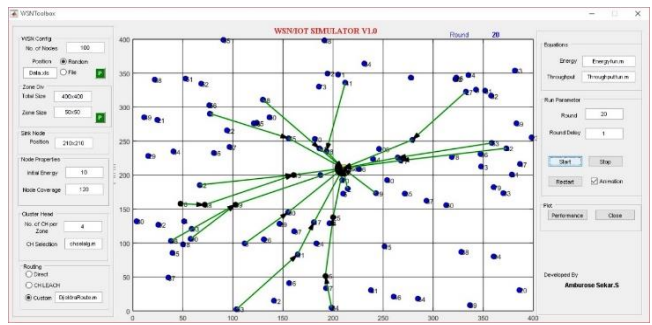

Fig 4: The configuration Parameters

The researchers in this paper, chose to use the IEEE 802.11s protocol mainly because it allows the user to model communication between multiple Wireless Local Area Network (WLAN) nodes containing medium access control (MAC) layer and PHY. The low-data rate Wireless Personal Area Network (WPAN) is used by many Internet of Things (IoT) protocols, such as ZigBee. MATLAB Simulink communication block set was used to build a complete WSN system. The simulation procedure includes building the hardware architecture of the transmitting nodes, modeling both the communication channel and the receiving master node architecture. IEEE 802.11s protocol was used because it brings interworking and security by defining the secure password-based authentication and key establishment protocol.

### 5. SIMULATION RESULTS AND ANALYSES

A simulation is beneficial as it allows the user to observe the efficiency of the proposed scheme before implementing the algorithm in a real network environment. A network simulator allows researchers to review the performance of the proposed algorithm under various network conditions. The simulator can be modified to archive better results for the required analysis. This section presents and analyses the outcome from the used simulation to evaluate the effectiveness of the User-Based authentication scheme from a variety of network configurations. Xgraph is used to present the evaluated MATLAB simulation results. Simulation results were obtained from 10 simulations.

### 5.1 Experimental Evaluation

To measure the effectiveness of the User-Based authentication and DoS mitigation scheme to protect WBSNs, experiments were provided. The MATLAB simulation version R2020a was used, 100 random nodes were deployed which can run 10 or more rounds in the IEEE 802.11s protocol and each simulation was performed for the 60s to get stable output statistic. The minimum distance between nodes is 0.6 cm. The approach uses the IEEE 802.11s protocol for allowing the communication between multiple wireless networks between nodes. RC4 encryption was used for the encryption/decryption process, where different length keys are known to a receiver.

### 5.2 Simulation Results

A simulation scenario was applied to perform the simulations as explained in Section 4.1 to monitor the performance of a User-Based authentication scheme with Cryptography-based techniques and Biometric-based authentication algorithm. The simulation results presented in this paper have been gathered from an average of 10 simulations. The performance metrics analysed in the simulations focus on the below parameters:

1) Network throughput: is the rate at which several bits are transferred across communication channels. It measures how fast data is transferred from the source to the destination.

2) Number of packet loss: Occurs when one or more packets of data responsible for transmitting data are failing to reach their destination.

3) Encryption/Decryption time: The encryption parameter measures the time taken when executing the encryption process, where an algorithm that takes less encryption time is considered efficient. Decryption is responsible for the amount of time the algorithm takes to decrypt the key.

This paper compared the performance of the proposed scheme with Cryptography-based and Biometric-based techniques based on user authentication. Biometric authentication algorithm was chosen because it provides a digital identity that is stored on network storage. Therefore, the process generates some time cost and consumes computational resources. In many cases, identification algorithms based on interactive proof are applied to authentication. The digital identity must be encrypted before sending, and the secret keys must also be distributed. When Biometric authentication is used in a network the user-authentication is maintained.

Cryptographic techniques are compared with the proposed algorithm because this technique uses the combination of asymmetric and symmetric cryptography when both parties have to agree on a secret key. After that, each message is encrypted with that same key, transmitted, and decrypted with the same key. However, Secure key distribution is the biggest problem when using symmetric cryptography. This paper compared the proposed authentication scheme with the Cryptographic techniques because both utilize keys to encrypt/decrypt messages. However, the proposed algorithm considered the authentication of the user by integrating ECC and RC4 in order to ensure that encryption/decryption is faster so that the network is not attacked while sending data through the transmission channel.

### 5.3.1 Effects of Network throughput

The effects of the network throughput capacity limitations on the performance of the three algorithms are presented as shown in Fig. 5. The performance of the three algorithms was analyzed when the gateway throughput capacity limitation differs from 2 to 18 while the delay limitation is set at 10 and the relay load constraint is relaxed.

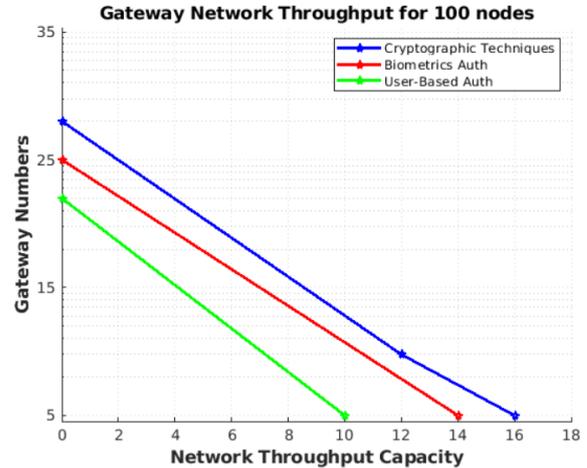

Fig 5: Comparison of the impacts of the gateway constraint on the performance of the three algorithms

Average network throughput for the three algorithms were monitored and compared under various network capacities being the volume of data sent, received, and speed at which data is transferred during the user authentication process. Biometrics authentication algorithm and Cryptographic techniques scheme are close to each other and achieved a higher gateway convergence as compared to User-Based authentication scheme.

### 5.3.2 Number of packet loss

The parameter measures the probability of packet loss in the communication network. In this case, the program has a sufficient duration of 50 seconds. The program transfers 20 packets, this allows the researcher to monitor and gather information about the number of packet loss, sent, and received during the authentication process.

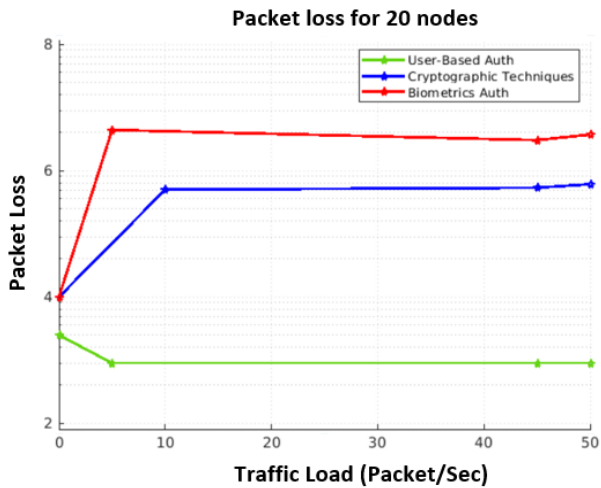

Fig 6: Average packet loss during the authentication process.

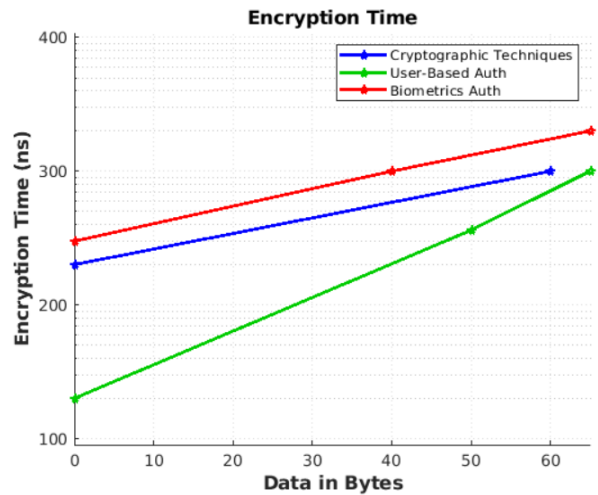

Fig 7: Encryption time versus byte of data

Simulation is repeated 10 times to obtain the average probability of packet loss for more accurate results. The proposed scheme produced 30% at the time interval of 50 seconds where only 2.5% of packets were lost. The proposed scheme outperforms both the Biometric authentication algorithm and Cryptographic techniques because the scheme uses ECC which is known to be fast, having a benefit in processing time because of its smaller keys. Hence reduced the number of packet loss. While the other two algorithms produced 40% at the time of 50 intervals where more than 50% of packets were lost.

*5.3.3 Effects of Encryption/Decryption time*

This parameter computes the time required for the execution of the User-Based authentication scheme for encryption and decryption. The algorithm that takes less encryption and decryption time is considered efficient. The User-Based authentication scheme is tested for different lengths of the password. When the password length was 10 bytes, the User-Based authentication scheme took 160 ns to encrypt as shown in Fig. 7.

Using different lengths of the user password during the authentication process is done to get the amount of time it takes for each password to encrypt if the key is known to a receiver. User-Based authentication scheme takes less time to encrypt as compared to Cryptographic Techniques and Biometrics authentication algorithm. Cryptographic Techniques and Biometrics authentication algorithms took a long to encrypt data. When the password length was 10 bytes, the Cryptographic Techniques scheme took 220ns to encrypt data while the Biometrics authentication algorithm encryption time was higher.

The decryption parameter calculates the time required to convert the received ciphertext to plaintext. The same process performed for encryption was followed where a User-Based authentication scheme was evaluated using different lengths of passwords with an aim of analysing the time it takes for each data to decrypt if the key is known to the receiver. Multiple data lengths versus the amount of time it takes to decrypt is shown in Fig. 8. Biometric authentication and Cryptograph techniques took more time to decrypt the data as compared to the User-Based authentication scheme.

The encryption and decryption time of the three schemes is shown in Fig. 7 and Fig. 8. The results were calculated based on the amount of data versus the number of bytes. User-Based authentication scheme provides data security with less encryption and decryption time. While the Biometric authentication and Cryptograph techniques took more time to decrypt the data provided as compared to the User-Based authentication scheme.

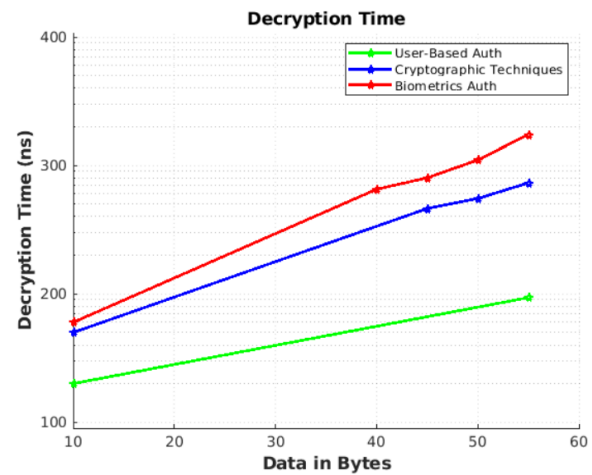

Fig 8: Decryption time versus byte of data.

6. CONCLUSION

The paper presented the design of the User-Based authentication and DoS mitigation algorithm by integrating ECC and RC4 in order to improve security in WBSNs. The effectiveness of the proposed User-Based authentication and DoS mitigation algorithm in WBSNs was evaluated against the Biometric authentication algorithm and Cryptographic techniques. The analysed results were presented based on the following network parameters namely network throughput, packet loss, encryption, and decryption time. According to the presented simulation results, the proposed algorithm outperformed both Biometric authentication algorithm and Cryptographic techniques in all mentioned network parameters.

In this paper, it is assumed that the computation time is not maintained during the user authentication process as the focus was on mitigating DoS attacks. Therefore, in the future, the computation time which includes key generation speed can be investigated to ensure that the security of WBSNs is maintained in all aspects and evaluating the performance of the User-Based Authentication algorithm in more complex network systems can be investigated.

ACKNOWLEDGMENT

The authors would like to thank the Tshwane University of Technology for financial support. The authors declare that there is no conflict of interest regarding the publication of this paper.